\documentclass{iau}
\usepackage{graphicx}

\title[Jets in AGN at extremely high redshifts] 
{Jets in AGN at extremely high redshifts
}

\author[L.I. Gurvits et al.]  
{
Leonid I. Gurvits$^{1,2}$, 
S\'{a}ndor Frey$^3$
 \and
Zsolt Paragi$^1$
}

\affiliation{
$^1$Joint Institute for VLBI in Europe, P.O. Box 2, 7990 AA Dwingeloo, The Netherlands \\ 
email: {\tt lgurvits@jive.nl, zparagi@jive.nl} \\
[\affilskip]
$^2$Dept. of Astrodynamics \& Space Missions, Delft University of Technology, \\ 
2629 HS Delft, Delft, The Netherlands \\
[\affilskip]
$^3$F\"{O}MI Satellite Geodetic Observatory, P.O. Box 585, H-1592 Budapest, Hungary \\
email: {\tt frey@sgo.fomi.hu}
}

\pubyear{2014}
\volume{313} 
\pagerange{xx--xx}
\jname{Extragalactic jets from every angle}
\editors{F. Massaro, C.C. Cheung, E. Lopez, A. Siemiginowska, eds.}

\begin{document}

\maketitle

\begin{figure}
\centering
  \includegraphics[bb=70  75 520 720, width=35mm, angle=270, clip=]{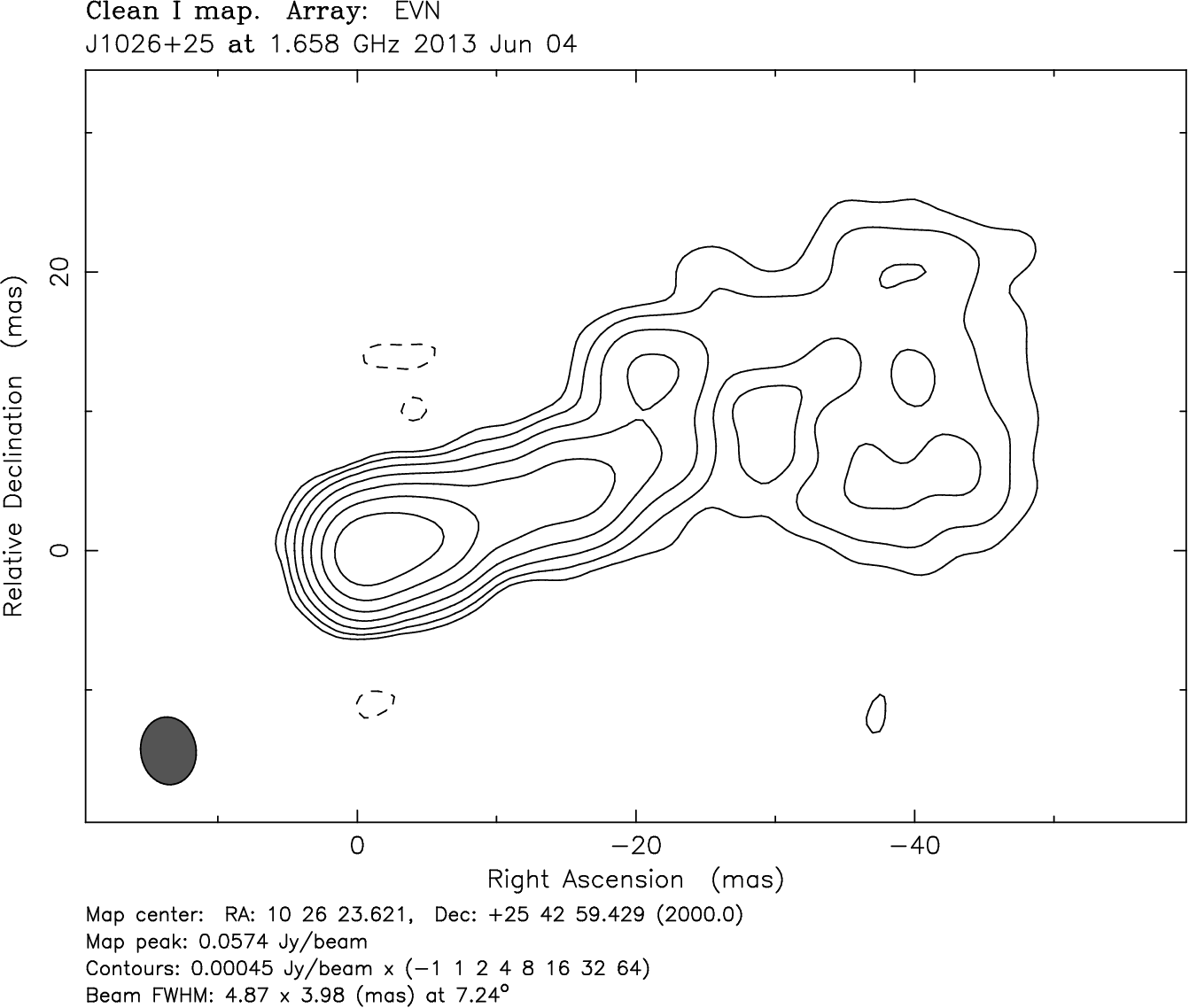}
  \includegraphics[bb=70 220 520 581, width=35mm, angle=270, clip=]{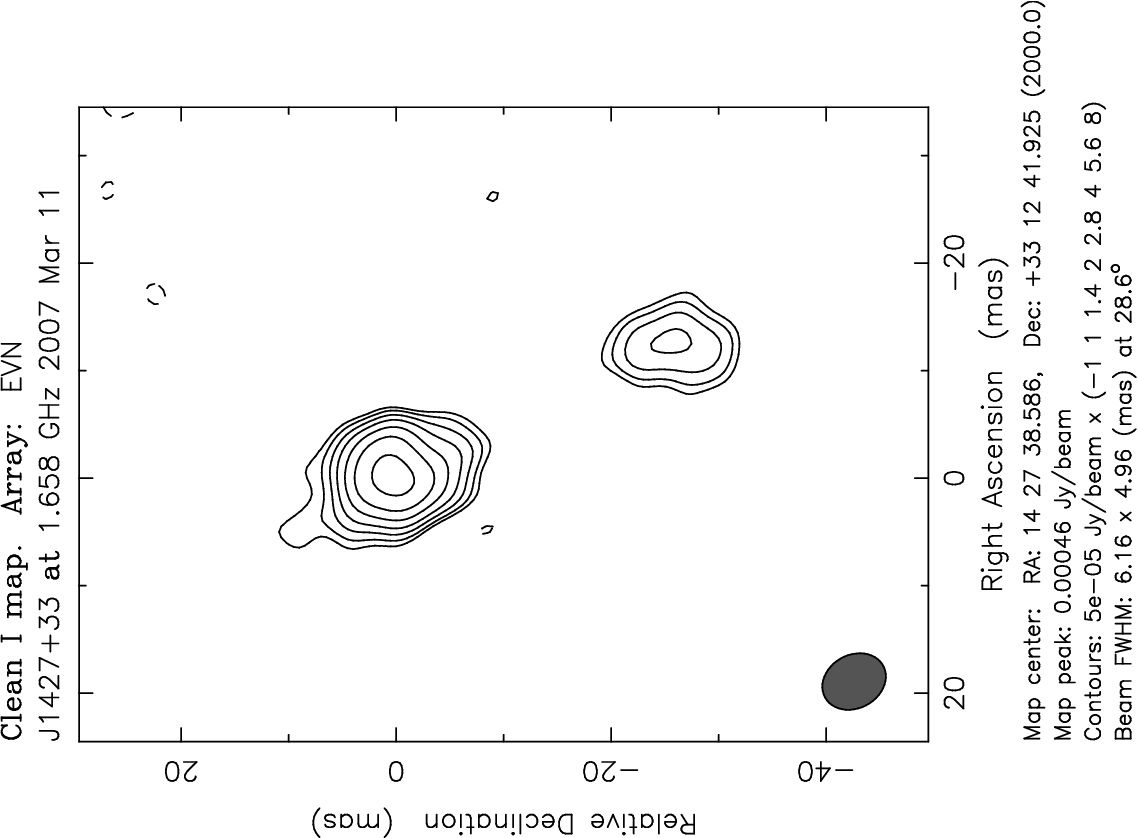}
  \includegraphics[bb=70 169 520 626, width=35mm, angle=270, clip=]{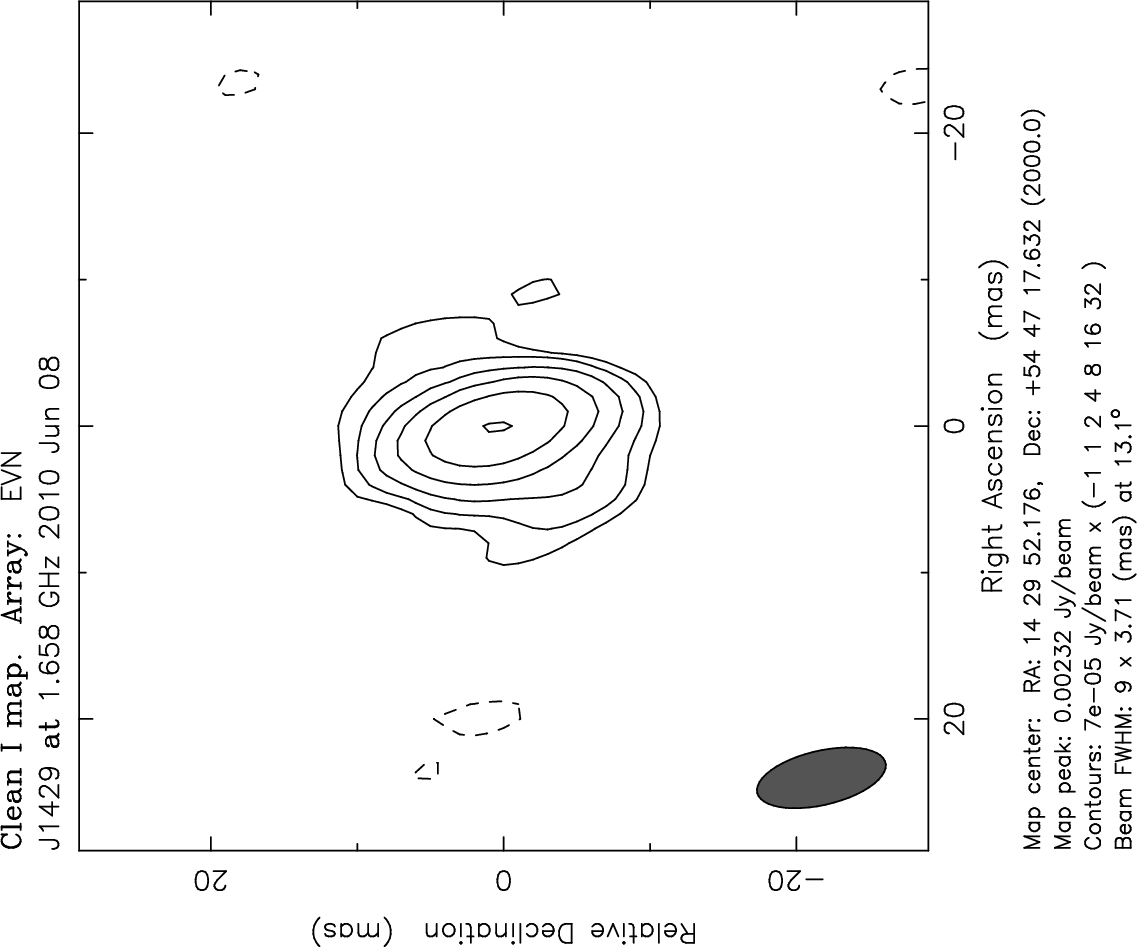}
  \caption{
VLBI images at 1.6 GHz of three very high-redshift quasars J1026$+$2542 ($z=5.266$, left), J1427$+$3312 ($z=6.12$, middle), and J1429$+$5447 ($z=6.21$, right). Parameters of images are given in \cite[Frey \etal \, (2015)]{Frey-2015}, \cite[Frey \etal \, (2008)]{Frey-2008}, and \cite[Frey \etal \, (2011)]{Frey-2011}, respectively.}
  \label{fig1}
\end{figure}

The jet phenomenon is a trademark of active galactic nuclei (AGN). In most general terms, the current understanding of this phenomenon explains the jet appearance by effects of relativistic plasma physics. The fundamental source of energy that feeds the plasma flow is believed to be the gravitational field of a central supermassive black hole. While the mechanism of energy transfer and a multitude of effects controlling the plasma flow are yet to be understood, major properties of jets are strikingly similar in a broad range of scales from stellar to galactic. They are supposed to be controlled by a limited number of physical parameters, such as the mass of a central black hole and its spin, magnetic field induction and accretion rate. In a very simplified sense, these parameters define the formation of a typical core--jet structure observed at radio wavelengths in the region of the innermost central tens of parsecs in AGN. These core--jet structures are studied in the radio domain by Very Long Baseline Interferometry (VLBI) with milli- and sub-milliarcsecond angular resolution. Such structures are detectable at a broad range of redshifts. If observed at a fixed wavelength, a typical core--jet AGN morphology  would appear as having a steep-spectrum jet fading away with the increasing redshift while a flat-spectrum core becoming more dominant. If core--jet AGN constitute the same population of objects throughout the redshift space, the apparent ``prominence'' of jets at higher redshifts must decrease \cite[(Gurvits 1999)]{Gurvits-1999}: well pronounced jets at high $z$ must appear less frequent than at low $z$. 

Owing to recent extensive optical surveys, a number of AGN identified at extremely high redshifts has grown significantly over the last decade. However, among a hundred or so identified AGN at $z>5$, only a handful have radio flux densities at mJy or higher levels, including just three quasars at $z>5.7$: J0836$+$0054, $z = 5.77$,  J1427$+$3312, $z = 6.12$, and J1429$+$5447, $z = 6.21$. They all have been imaged with VLBI \cite[(Frey \etal \, 2011, and references therein)]{Frey-2011}, with the latter two shown in Fig.\,\ref{fig1} (middle and right). 

Studies of compact structures in inner jets of quasars address two objectives. First, comparison of structural and kinematic properties of AGN jets across the redshift space allows us to search for possible differences in physical properties of these objects at widely separated cosmological epochs and/or reveal morphological properties indicative of intrinsic evolution. Second, if the radio structures of inner jets in AGN prove to be intrinsically ``standard'' or described by a limited number of physical parameters, their redshift-dependent appearance might serve as a means of independent verification of the cosmological model. Similar studies have been conducted in the 1990s on much smaller samples of VLBI-imaged AGN in a more narrow range of redshifts. However, the accessible highest redshift has grown from about 3.8 in the 1990s to about 6.2 at present. The number of sources available for the redshift-dependent studies of VLBI structures has also grown by nearly an order of magnitude.

A prevalence of gigahertz peaked-spectrum (GPS) and compact steep-spectrum (CSS) sources, which are believed to be young (i.e., being at earlier stages of their evolution) or ``frustrated'' (i.e., confined by the dense interstellar medium) in AGN at high redshifts has been predicted by \cite[Savage \& Peterson (1983)]{Savage-1983} and analysed by \cite[O'Dea (1998)]{O'Dea-1998}. Their studies addressed samples of quasars at considerably lower redshifts than accessible now. Indeed, the spectra and morphologies identified in the object J1427$+$3312 at $z=6.12$ \cite[(Frey \etal \, 2008)]{Frey-2008} and suspected in J1429$+$5447 at $z=6.21$ \cite[(Frey \etal \, 2011)]{Frey-2011} are consistent with what would be expected in a young GPS/CSS source.

However, within the statistically unconvincing sample of VLBI-imaged extremely high-redshift sources, appearance of canonical core--jet morphology is not uncommon either. There is a hint on such the morphology dominated by a core in one of the weakest among VLBI-imaged high-$z$ sources, the quasar J0836$+$0054 at $z=5.774$ \cite[(Frey \etal \, 2005)]{Frey-2005}.

A more obvious case of the core--jet morphology is found in the quasar J1026$+$2542 at $z=5.266$ (Fig.\,\ref{fig1}, left). VLBI images of this source obtained at two epochs separated by $\sim$7 years (\cite[Helmboldt \etal , 2007,]{Helm-2007} and \cite[Frey \etal , 2013]{Frey-2013}) allow us to estimate for the first time the proper motion in a jet at such a high redshift. For the four jet structural components, the proper motion estimates are in the range from 0.03 to 0.11 mas/yr consistent with the Lorentz factor $\Gamma \simeq 12.5$ \cite[(Frey \etal \, 2015)]{Frey-2015}.  We note that proper motion measurements in very high-redshift sources require longer time baselines for a given angular resolution due to the time dilation proportional to $(1+z)$. 

The accumulating new observational data on high-redshift radio quasars, including those shown here, will soon warrant a new look onto the VLBI structures of inner AGN jets as cosmological probes.

{\it Acknowledgements:} SF and ZP thank for the support of the Hungarian Scientific Research Fund (OTKA K104539).

\end{document}